\documentclass[onecolumn,sort&compress,numbers]{els-mrw} 

\usepackage{amsmath,amssymb,amsfonts,amsthm,makeidx,graphicx}
\usepackage{txfonts}
\usepackage{helvet}
\usepackage{hyperref}
\usepackage{cancel}

\renewcommand{\Re}{\text{Re}\,}
\renewcommand{\Im}{\text{Im}\,}
\newcommand{\disc}{\text{disc}\,}
\newcommand{\diff}{\text{d}} 
\renewcommand{\vec}[1]{\ensuremath{\mathbf{#1}}}
\newcommand{\F}{\mathcal{F}}
\newcommand{\Order}{\mathcal{O}}

\begin{document}

\chapter{Dispersion relations: foundations}\label{chap1}

\author[1]{Bastian Kubis}%

\address[1]{Helmholtz-Institut für Strahlen- und Kernphysik (Theorie) and Bethe Center for Theoretical Physics, Universität Bonn, \mbox{53115 Bonn}, Germany}

\articletag{Chapter Article tagline: update of previous edition, reprint.}

\maketitle

\begin{abstract}[Abstract]
We give a pedagogical introduction to the founding ideas of dispersion relations in particle physics.  Starting from elementary mechanical systems, we show how the physical principle of causality is closely related to the mathematical property of analyticity, and how both are implemented in quantum mechanical scattering theory.  We present a personal selection of elementary applications such as the relation between hadronic production amplitudes or form factors to scattering, and the extraction of resonance properties on unphysical Riemann sheets.  More advanced topics such as Roy equations for pion--pion scattering and dispersion relations for three-body decays are briefly touched upon.
\end{abstract}

\begin{keywords}
 	analyticity and causality\sep unitarity and probability conservation \sep form factors \sep resonance poles 
\end{keywords}

\section*{Objectives}
\begin{itemize}
	\item How can the mathematical concept of analyticity be understood as a consequence of the physical requirement of causality?
	\item What are the analytic properties of quantum mechanical scattering amplitudes?
	\item How can dispersion relations for scattering amplitudes or form factors be derived as a consequence of analyticity and unitarity?
	\item How can resonance properties be rigorously determined from poles on unphysical Riemann sheets?
\end{itemize}

\section{Introduction}\label{intro}

Dispersion relations are a powerful tool for analyzing quantum mechanical scattering, decay, or production amplitudes.  Their main strength lies in their foundation in very general principles---causality, first of all, which translates into the mathematical property of analyticity, as we will discuss in some detail---that physicists take for granted quite in general.  As such, dispersion relations can be applied in many different contexts in elementary particle physics; but from the perspective of someone interested specifically in the strong interactions at low energies, where the underlying quantum field theory, quantum chromodynamics, is inaccessible to perturbation theory, one of its major strengths is that these principles apply beyond a perturbative setting.  

The consequences of causality/analyticity are often combined with other general principles.  One central additional idea is unitarity as the consequence of probability conservation; in relativistic theories, crossing symmetry is another such principle.  Furthermore, depending on the precise context, also other symmetries such as the discrete ones of parity, time reversal, and charge conjugation can be taken into account, or flavor symmetries such as isospin.  Finally, the combination of dispersion relations with ideas from effective field theories often proves to be particularly useful.

Dispersion relations can be used for consistency checks and constraints applied to quantum mechanical amplitudes (and the techniques to implement these may vary widely); but they also allow us to derive central results closely related to analytic properties, e.g., of resonance poles.  

The following introduction to the foundations of dispersion relations is largely informal and devoid of mathematical rigor; we will point to some more in-depth treatments in the end.  We start by recalling some basics from complex analysis in Sec.~\ref{sec:complexanalysis}, and then illustrate the connection between analyticity and causality based on an example from classical mechanics in Sec.~\ref{sec:analyticity}.  In Sec.~\ref{sec:NRscattering}, we show that non-relativistic quantum mechanical scattering amplitudes are analytic, and derive a first dispersion relation.  We then apply this in the relativistic, quantum field theoretical context, first for a Feynman diagram from perturbation theory in Sec.~\ref{sec:Feynman-diagrams}, subsequently for a non-perturbative object, the pion form factor, in Sec.~\ref{sec:PionFF}.  Subsequently, we discuss how to extract resonance properties rigorously in Sec.~\ref{sec:resonances}.  Finally, we comment on more advanced topics, such as the practical application of dispersion relations to hadron scattering in Sec.~\ref{sec:scattering}, and to three-body decays in Sec.~\ref{sec:decays}, before we conclude in Sec.~\ref{sec:conclusions}.

\section{Some mathematical basics from complex analysis}\label{sec:complexanalysis}

As the use of complex analysis will be central for much that we will discuss in the following, let us very briefly recall some very basic results that will prove useful later.

\subparagraph{Cauchy--Riemann condition}

Let $f(z)$ be a complex function, $z = x + i y$, and $f(x,y) = u(x,y) + i v(x,y)$.  
$f$ is complex differentiable if and only if $u$ and $v$ are real differentiable and fulfill the Cauchy--Riemann condition
\begin{align}
\frac{\partial u}{\partial x} = \frac{\partial v}{\partial y}, 
\qquad 
\frac{\partial u}{\partial y} = -\frac{\partial v}{\partial x}.
\end{align}
$f(z)$ is \textit{analytic} at $z_0$ if $f'(z)$ exists at $z_0$ and in a neighborhood of $z_0$.  $f(z)$ is then also called holomorphic.

\subparagraph{Cauchy integral theorem}

If $f(z)$ is analytic within a domain $D$, then
\begin{align}
\oint_C f(z)\, \diff z = 0,
\end{align}
where $C$ denotes an arbitrary simple (singly connected) closed contour within $D$.
As a very simple example, consider $f(z) = z^n$, $n \in \mathbb{N}_0$, and a clockwise contour with $|z|=R$. Then
\begin{align}
\oint f(z)\, \diff z = \oint z^n \, \diff z
= i R^{n+1} \int_0^{2\pi} \diff\theta \, e^{i(n+1)\theta} 
= \frac{R^{n+1}}{n+1} \Bigl( e^{i(n+1)\theta} \Bigr) \Big|_0^{2\pi} 
= 0,
\end{align}
as expected.
In fact, the above example is trivially generalized to $f(z) = z^n$, $n \in \mathbb{Z}$, i.e., to functions that are non-analytic at $z=0$ for $n < 0$, with the exception of $n=-1$ for which the derivation obviously does not hold:  
\begin{align}
\oint \frac{\diff z}{z} = i R^0 \int_0^{2\pi} \diff\theta \, e^{0} = 2\pi i.
\end{align}
This leads us to the

\subparagraph{Residue theorem}
Let $f(z)$ be meromorphic, i.e., holomorphic except for isolated poles at $z_1, \ldots, z_n$.  
Then
\begin{align}
\oint_C f(z)\, \diff z = 2\pi i \sum_{j=1}^n \operatorname{Res}_{z_j} f(z),
\end{align}
where $C$ is a simple closed path (with winding number $+1$).  
The residues are defined by
\begin{align}
\operatorname{Res}_{z_0} f(z) = \frac{1}{2\pi i} \oint_{C_0} f(z)\, \diff z,
\end{align}
where $C_0$ encloses only the pole at $z_0$.  
For simple poles,
$\operatorname{Res}_{z_0} f(z) = \lim_{z \to z_0} (z - z_0) f(z)$,
i.e., near $z=z_0$ we can expand
\begin{align}
f(z) = \frac{\operatorname{Res}_{z_0} f(z)}{z-z_0} + \sum_{n=0}^\infty a_n (z-z_0)^n.
\end{align}

\subparagraph{Cauchy integral formula}

If $f(z)$ is analytic inside a domain $D$, then
\begin{align}
\frac{1}{2\pi i} \oint_{\partial D} \frac{f(z)}{z-z_0}\, \diff z = f(z_0).
\end{align}

Analytic or holomorphic functions are in fact infinitely differentiable.  
The relevance for physics (as we will see) is that functions at one $z_0$ can be related to an integral over different $z$.  We will transfer this finding to quantum mechanical (or quantum field theoretical) amplitudes.
What will be important to show next is that relevant physical amplitudes are indeed analytic.

\section{Analyticity as a consequence of causality}\label{sec:analyticity}

To illustrate the connection between the mathematical property of \textit{analyticity} and the physical requirement \textit{causality}, we follow Ref.~\cite{Nussenzveig:1972tcd} and begin with an example from classical mechanics: the forced damped harmonic oscillator.  It is described by the differential equation
\begin{align}
\ddot x(t) + 2\gamma \dot x(t) + \omega_0^2 x(t) = f(t) , \label{eq:HO}
\end{align}
where $\gamma>0$ is the damping factor, $\omega_0$ the eigenfrequency of the free, undamped oscillation, and $f(t)$ the external force. The free solutions for $f(t) \equiv 0$ are easily found to be 
\begin{align}
x_{\text{free}}(t) = a\,e^{-i\omega_1t} + b\,e^{-i\omega_2t} , \qquad \omega_{1/2} = \pm\sqrt{\omega_0^2-\gamma^2} - i \gamma ,
\end{align}
where $\omega_{1/2}$ are the solutions of
\begin{align}
\omega^2+2i\gamma\omega-\omega_0^2 = (\omega-\omega_1)(\omega-\omega_2) = 0 .
\end{align}
The eigenfrequencies of the system are complex, with $\Im \omega_{1/2} < 0$, such that the solutions are damped by a factor $e^{-\gamma t}$.  We see that the positivity of $\gamma$ is a central requirement, as otherwise the oscillations would grow infinitely, which would clearly be unphysical.  

\begin{figure}[t]
	\centering \hspace*{1cm}
	\includegraphics[width=.4\textwidth]{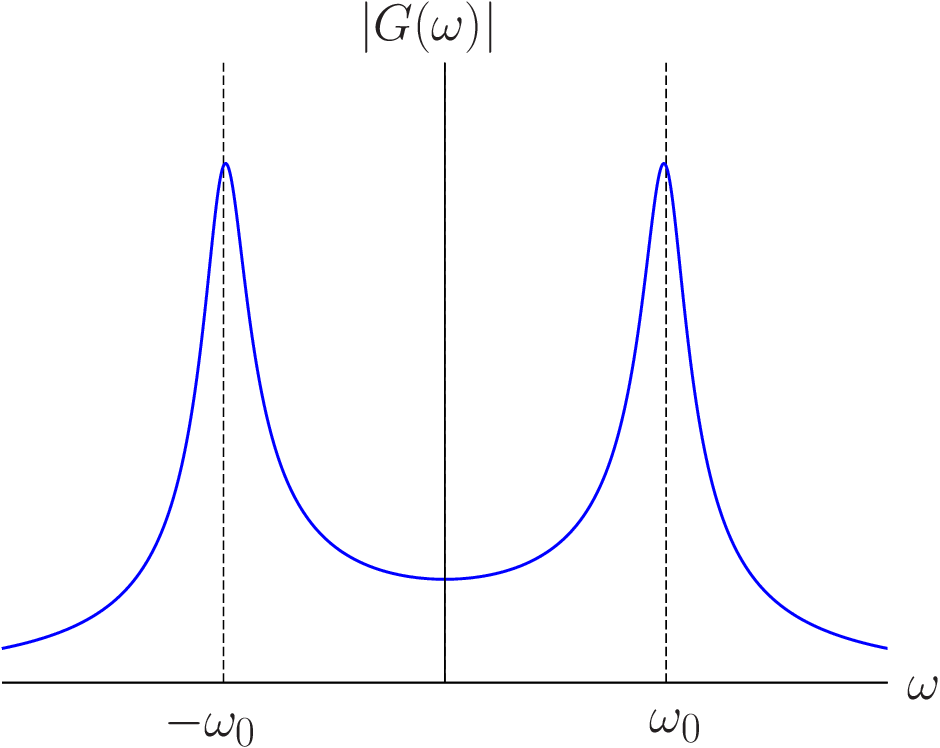} \hfill
    \includegraphics[width=.4\textwidth]{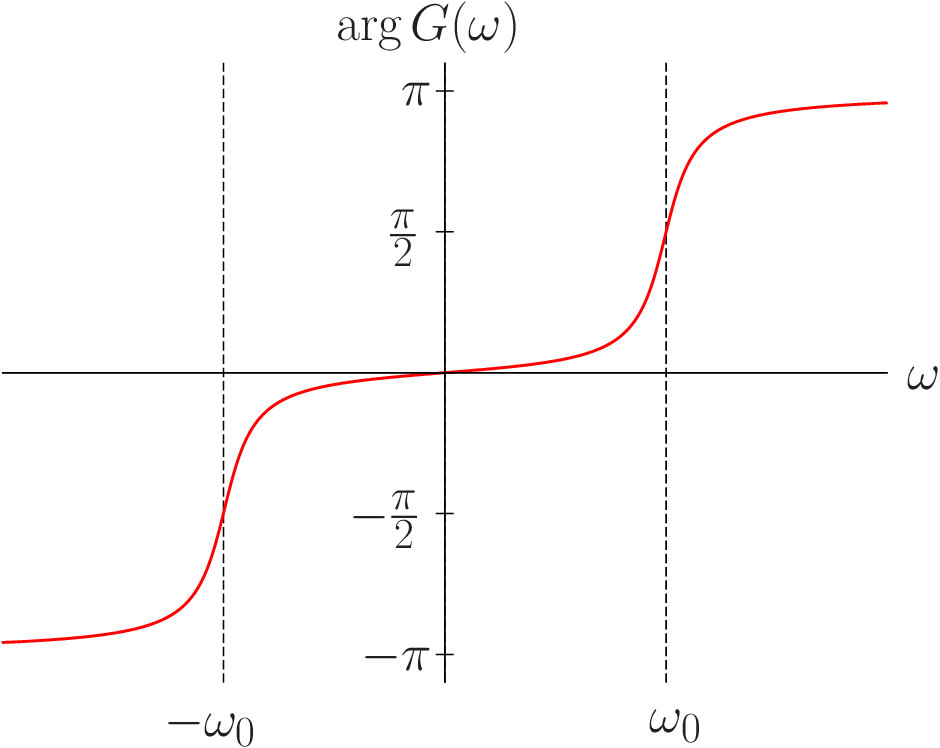} \hspace*{1cm}
	\caption{Modulus (left) and phase (right) of the Green's function $G(\omega)$, with resonant behavior at $\omega \approx \pm \omega_0$.}
	\label{fig:Greens}
\end{figure}

We begin the analysis of the forced oscillator with a harmonic force, $f(t) = F_\omega e^{-i\omega t}$, which leads to a special solution
\begin{align}
x(t) =  G(\omega) F_\omega e^{-i\omega t} , \qquad \text{where} \quad G(\omega) = -\frac{1}{(\omega-\omega_1)(\omega-\omega_2)}
\label{eq:HOresponse}
\end{align}
is the response function.  For weak damping $\gamma \ll \omega_0$, $G(\omega)$ shows \textit{resonant} behavior at $\omega\approx \pm \sqrt{\omega_0^2-\gamma^2} \approx \pm \omega_0$, characterized by peaks in its modulus and a rapid change in its phase, cf.\ Fig.~\ref{fig:Greens}.

To solve Eq.~\eqref{eq:HO} for arbitrary forces $f(t)$, we use the Fourier transform $F(\omega)$, 
\begin{align}
    f(t) = \frac{1}{2\pi} \int_{-\infty}^\infty \diff\omega \, F(\omega) \, e^{-i\omega t} .
\end{align}
Since the system is linear, the solution is a superposition of solutions of the form~\eqref{eq:HOresponse},
\begin{align}
    x(t) = \frac{1}{2\pi} \int_{-\infty}^\infty \diff\omega \, G(\omega) F(\omega) e^{-i\omega t} + \text{homogeneous solution}.
\end{align}
By inserting the inverse Fourier transform $F(\omega) = \int_{-\infty}^\infty \diff t' f(t') e^{i\omega t'}$, we can rewrite this as
\begin{align}
    x(t) = \int_{-\infty}^\infty \diff t' g\left(t-t'\right) f\left(t'\right) , \qquad g\left(\tau\right) = \frac{1}{2\pi} \int_{-\infty}^\infty \diff\omega\, G(\omega)\,e^{-i\omega\tau} . \label{eq:Greens}
\end{align}
The \textit{Green's function} $g(\tau)$ describes the propagation of an instantaneous force $f\left(t'\right) = \delta\left(t'\right)$, with $\tau = t-t'$.  

\begin{figure}[t]
	\centering
	\includegraphics[width=.45\textwidth]{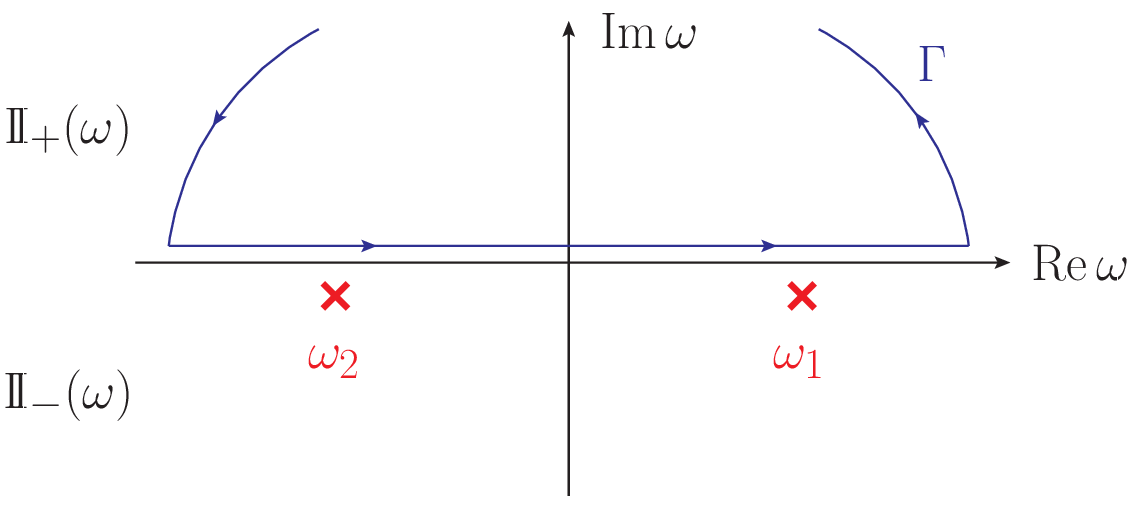}
	\caption{Integration in the complex $\omega$ plane along the contour $\Gamma$ (blue).  Both poles $\omega_{1/2}$ (red crosses) lie in the lower half plane $\mathbb{I}_-(\omega)$, while the integration contour is closed in the upper half plane $\mathbb{I}_+(\omega)$ where the Green's function is analytic.}
	\label{fig:frequency-plane}
\end{figure}

The requirement of \textit{causality} now implies that 
\begin{align}
    g(\tau<0) = 0,
\end{align}
since otherwise, a force from the future $t'>t$ would propagate to the past $t$.  Let us check this property for the explicit form of $g(\tau)$ in Eq.~\eqref{eq:Greens}.  We note that the response function $G(\omega)$ in Eq.~\eqref{eq:HOresponse} is analytic except for the two poles that lie in the lower half of the complex $\omega$-plane $\mathbb{I}_-(\omega)$, as $\gamma>0$.  As a result, we can close the integration contour in the upper half of the complex plane $\mathbb{I}_+(\omega)$, cf.\ Fig.~\ref{fig:frequency-plane}, 
and Cauchy's theorem implies
\begin{align}
\frac{1}{2\pi} \oint_\Gamma \diff\omega \, G(\omega) e^{-i\omega\tau} = 0.
\end{align}
For $\tau<0$, the integral over the complex arc vanishes for $|\omega| \to \infty$, since $|G(\omega)| < 1/|\omega^2-\omega_0^2|$, and $|e^{-i\omega\tau}| = e^{+\Im(\omega)\tau}$, which is exponentially suppressed for negative $\tau$.  As a result, 
\begin{align}
    g(\tau<0) = \frac{1}{2\pi} \int_{-\infty}^\infty \diff\omega\, G(\omega)e^{-i\omega\tau} = 0,
\end{align}
as required.  We therefore conclude
\begin{align}
    G(\omega) \quad \text{is analytic in} \quad \mathbb{I}_+(\omega) \qquad \Leftrightarrow \qquad \text{causality} \quad g(\tau<0) = 0 .
\end{align}

In general, if we assume a system with ``input'' $f(t)$ and ``output'' $x(t)$ that
\begin{enumerate}
\item is \emph{linear}, $x(t) = \int_{-\infty}^\infty \diff t' \, g\left(t,t'\right) f\left(t'\right)$;
\item is \textit{time-translation invariant}, $g\left(t,t'\right) = g\left(t-t'\right)$;
\item is \emph{causal}, $g\left(t-t'\right) = 0$ for $t'>t$;
\end{enumerate}
then the Fourier transform
\begin{align}
G(\omega) = \int_{-\infty}^\infty \diff\tau\, g(\tau) e^{i\omega\tau} = \int_{0}^\infty \diff\tau\, g(\tau) e^{i\omega\tau} \label{eq:analyticG}
\end{align}
is analytic in $\mathbb{I}_+(\omega)$.

We note that if the integral in Eq.~\eqref{eq:analyticG} converges for $\omega\in\mathbb{R}$, it does so even better for $\omega\in \mathbb{I}_+(\omega)$; this would, however, not be true if the integral had support from $\tau<0$.  A function as $G(\omega)$ in Eq.~\eqref{eq:analyticG} is also called a \textit{causal transform}.

\section{Scattering in non-relativistic quantum mechanics}\label{sec:NRscattering}
What we want to argue next is that (non-relativistic) quantum mechanical scattering amplitudes are causal transforms. They have an analytic continuation to complex arguments and a defined analytic structure. We here largely follow Ref.~\cite{Omnes:1971}.

We start from the Schrödinger equation with a spherically symmetric potential. The wave function is decomposed into spherical harmonics and the radial part according to
\begin{align}
\psi(\mathbf{r}) \to \frac{u_{\ell}(E, r)}{r} Y_{\ell}^{m} (\theta, \phi) ,
\end{align}
where $r = |\vec{r}|$.  The resulting radial Schrödinger equation for $u_\ell(E,r)$ reads 
\begin{align}
\left[ \frac{\diff^2}{\diff r^2} - \frac{\ell(\ell+1)}{r^2} - U(r) + k^2 \right] u_{\ell}(E, r) = 0 \quad \text{where} \quad
E = \frac{k^2}{2m}, \quad k = \pm\sqrt{2mE}, \quad U(r) = 2mV(r),
\label{eq:radialSEq}
\end{align}
and we assume $V(r \to \infty) \to 0$ \text{ faster than } ${1}/{r^2}$.
Asymptotically, for large distances, one has
\begin{align}
\left( \frac{\diff^2}{\diff r^2} + k^2 \right) u_{\ell}(E, r) = 0,
\qquad \text{resulting in} \qquad
u_{\ell}(E, r) \sim \phi_{\ell}^+(E) e^{-ikr} + \phi_{\ell}^-(E) e^{ikr},
\end{align}
where $\phi_{\ell}^+(E)$ and $\phi_{\ell}^-(E)$ correspond to incoming and outgoing waves, respectively.
For a vanishing potential, the regular solution is a spherical Bessel function
$u_{\ell}(r) \sim \sin(k r - {\ell\pi}/{2})$,
hence $\phi_{\ell}^- = - \phi_{\ell}^+$ in this limit.  The ratio of outgoing and incoming waves defines the \textit{$S$-matrix element}
\begin{align}
S_{\ell}(E) = (-1)^{\ell+1} \frac{\phi_{\ell}^-(E)}{\phi_{\ell}^+(E)} = e^{2i\delta_\ell(E)},
\end{align}
which can be written in terms of the phase shift $\delta_\ell(E)$ in the usual form. 
For simplicity, we largely confine ourselves to $S$-waves in the following, $\ell=0$, and drop the index, $\phi_0^\pm \to \phi^\pm$.
The time-dependent incoming and outgoing wave packets are given as
\begin{align}
\phi_{\text{in}}(r,t) &= \int_0^\infty \diff E \, \phi^+(E) e^{-i(k r + Et)} , \qquad
\phi_{\text{out}}(r,t) = \int_0^\infty \diff E \, \phi^-(E) e^{i(k r - Et)}. \label{eq:in-out}
\end{align}
To define the scattered wave, we write
\begin{align}
\phi_{\text{sc}}(r,t) = \int_0^\infty \diff E \left( \phi^-(E) - \phi^+(E) \right) e^{i(k r - Et)} = \int_0^\infty \diff E \left[ S(E) - 1 \right] \phi^+(E) e^{i(k r - Et)} . \label{eq:scatt}
\end{align}
We can interpret Eqs.~\eqref{eq:in-out} and \eqref{eq:scatt} as Fourier transforms with respect to time, with
\begin{align}
\phi_{\text{in}}(E) = \Theta(E) \phi^+(E) e^{-ikr} , \qquad
\phi_{\text{sc}}(E) = \Theta(E) \phi^+(E) \left[ S(E) - 1 \right] e^{ikr} .
\end{align}

We now recast $\phi_{\text{sc}}(r,t)$ in terms of a causal transform:
\begin{align}
\phi_{\text{sc}}(r,t) = \int_{-\infty}^\infty \diff t' \, g(r, t - t') \phi_{\text{in}}(r, t'), \qquad \text{where} \quad
g(r,\tau) = \int_0^\infty \diff E \, G(r, E) e^{-iE\tau}, \quad
G(r, E) = \frac{1}{2\pi} \left[ S(E) - 1 \right] e^{2i k r}.
\end{align}
If $ g(r, \tau < 0) = 0 $ due to causality, we have a causal transform and
$G(r, E)$  is analytic in  $\mathbb{I}_+(E)$.

Now recall that all the coefficients in the radial Schrödinger equation~\eqref{eq:radialSEq} are real. Hence, 
if $k^2$ has a solution $u(k^2)$ with $\phi^{\pm}(k^2)$, then $(k^2)^*$ has a solution $u(k^{2*})$ with $\phi^{\pm}(k^{2*}) = \big[\phi^{\pm}(k^2)\big]^*$. 
Therefore,
\begin{align}
S(E^*) = \big[S(E)\big]^* , \label{eq:Schwarz}
\end{align}
and the $S$-matrix element $S(E)$ in the lower half plane $\mathbb{I}_-(E)$ is determined from the upper half plane $\mathbb{I}_+(E)$, and is therefore analytic both in $\mathbb{I}_+(E)$ and in $\mathbb{I}_-(E)$. Equation~\eqref{eq:Schwarz} shows that $S(E)$ fulfills the Schwarz reflection principle.   

To describe scattering, we need both the momentum $k$ and the energy $E$; we chose the relation $k = +\sqrt{k^2} = +\sqrt{2mE}$, but there is another branch for $k = -\sqrt{k^2}$.
We have considered the $(+)$ branch so far and derived analyticity constraints there; however, as we will see below, these do not exist on the $(-)$ branch (where we will find resonance poles).
The usual convention is that, for real $k$, 
\begin{align}
S_{\text{physical}}(\Re k) = \lim_{\epsilon \to 0^+} S(\Re k + i\epsilon)
\end{align}
defines the \textit{physical} or first Riemann sheet.
However, we can show that the information on the second $(-)$ sheet is redundant with respect to the first one:
\begin{align}
\phi^{+}(k) = \phi^{-}(-k), \quad 
S_{\text{I}}(k) = -\frac{\phi^{-}(k)}{\phi^{+}(k)} , \quad \text{therefore} \quad
S_{\text{II}}(k) = S_{\text{I}}(-k) = -\frac{\phi^{+}(k)}{\phi^{-}(k)} = \frac{1}{S_{\text{I}}(k)} .
\end{align}
In other words, the $S$-matrix element on the second $k$-sheet $S_{\text{II}}$ is the \textit{inverse} of the $S$-matrix element on the first sheet $S_{\text{I}}$.

\subparagraph{Singularities}

We have seen that on the first sheet, $S(E)$ is analytic both for positive and for negative imaginary parts of the argument $E$; the only region where we have not excluded singularities is for real energies, $E \in \mathbb{R}$.  
Two kinds exist:
\begin{description}
\item[The ``physical'' or ``right-hand'' cut.]
Approaching the real axis from above and below, we observe
\begin{align}
\disc S(E) \equiv
S(E + i\epsilon) - S(E - i\epsilon) = S(E + i\epsilon) - S^*(E + i\epsilon) 
= 2i\, \Im S(E + i\epsilon) . \label{eq:disc-Im}
\end{align}
Thus, whenever $\Im S(E + i\epsilon) \neq 0$, there will be a \textit{discontinuity} in the function $S(E)$ on the real axis.
In general, $\Im S(E) \neq 0$ when there is a physically accessible state, which happens above threshold $E > 0$.  
Hence, there is a cut in $S(E)$ from threshold to $E = +\infty$.

\item[Bound states.]
We recall that the scattering wave function for large $r$ behaves according to
\begin{align}
u_\ell(r) \propto e^{-ikr} + S(k) e^{+ikr}, \label{eq:asymptotic-wave}
\end{align}
where the first term refers to the incoming wave and the second to the outgoing one, and only the ratio $S(k)$ (with a slight abuse of notation for convenience) of both is observable.
On the other hand, for a bound-state solution with negative energy
$
E = -{\kappa^2}/{2m} < 0 
$,
the wave function behaves asymptotically like a real exponential:
\begin{align}
u(r) \propto \cancel{A e^{\kappa r}} + B e^{-\kappa r}, \label{eq:bound-state}
\end{align}
where only the second term leads to a normalizable solution localized in space.  Such bound states exist for discrete values of $\kappa$. Equation~\eqref{eq:bound-state} is rather similar to Eq.~\eqref{eq:asymptotic-wave}, with $k \to i\kappa$, however, without the analog of an incoming wave:  
$
S(i\kappa) = -{\phi^{-}(i\kappa)}/{\phi^{+}(i\kappa)}  
$ is \textit{infinite} for a bound state.
Thus, $S(k)$ has a pole at $k = i\kappa$, or, equivalently, $S(E)$ has a pole at $E = -{\kappa^2}/{2m}$.  
\end{description}

\subparagraph{Dispersion relation}
\begin{figure}[t]
	\centering
	\includegraphics[width=.45\textwidth]{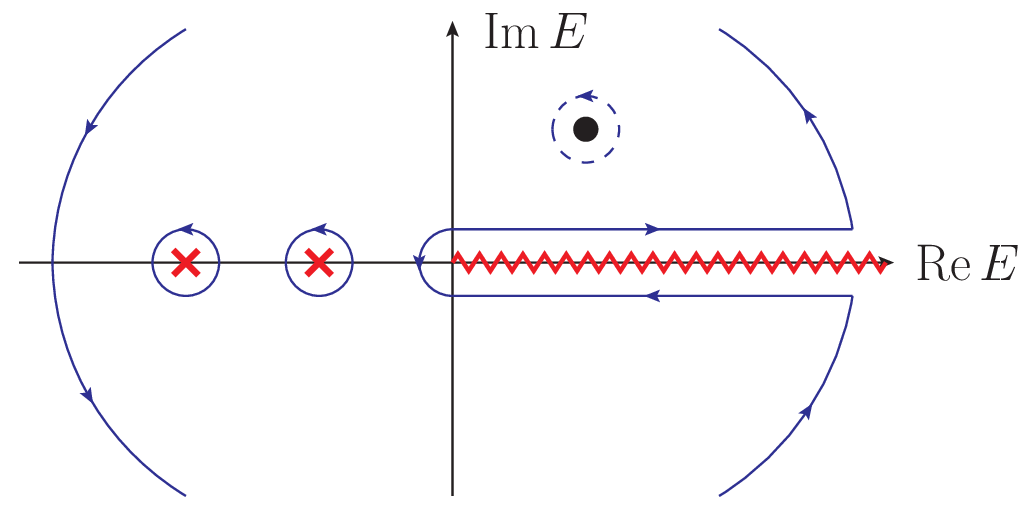}
	\caption{Analytic structure of $S(E)$, with a right-hand cut (red zig-zag line) for $E>0$ and bound states (red crosses) for $E<0$.  Extending the original contour of the Cauchy integral (blue dashed) towards infinity leads to the blue contour that wraps around both cut and poles.}
	\label{fig:S-matrix-analytic}
\end{figure}
The general analytic structure of $S(E)$ (for reasonable potentials) is therefore of the form as indicated in Fig.~\ref{fig:S-matrix-analytic}; which allows us to formulate our first dispersion relation.
We are mainly interested in
$E = E_0 + i\epsilon$, $\epsilon > 0$, for $E_0 \in \mathbb{R}^+$.
From Cauchy's theorem, 
\begin{align}
S(E) = \frac{1}{2\pi i} \oint \frac{S(E')\, \diff E'}{E' - E}.
\end{align}
If $S(E\to\infty) \to 0$ sufficiently fast, the large arc vanishes, yielding
\begin{align}
S(E) &= \frac{1}{2\pi i} \int_{0}^{+\infty} \frac{S(E' + i\epsilon)}{E' - E}\, \diff E'
+ \frac{1}{2\pi i} \int_{+\infty}^{0} \frac{S(E' - i\epsilon)}{E' - E}\, \diff E' 
+ \sum_{j} \frac{\text{Res}\, S(E_j)}{E_j-E} \notag \\
&= \frac{1}{2\pi i} \int_{0}^{+\infty} \frac{S(E' + i\epsilon)-S(E'-i\epsilon)}{E' - E}\, \diff E'
+ \sum_{j} \frac{\text{Res}\, S(E_j)}{E_j-E}
= \frac{1}{\pi} \int_{0}^{+\infty} \frac{\Im S(E' + i\epsilon)}{E' - E}\, \diff E'
+ \sum_{j} \frac{\text{Res}\, S(E_j)}{E_j-E}
, \label{eq:DR-step1}
\end{align}
where the sums in $j$ are meant to comprise all possible bound states, and we have used Eq.~\eqref{eq:disc-Im}.  
To approach the real axis $\epsilon \to 0$, recall the Sokhotski--Weierstrass formula
\begin{align}
\frac{1}{x - x_0 \mp i\epsilon} \xrightarrow{\epsilon \to 0} \frac{\mathcal{P}}{x - x_0} \pm i\,\pi \,\delta(x - x_0) , \label{eq:SokhWeier}
\end{align}
where $\mathcal{P}$ denotes the principal-value integral. 
Inserting this into Eq.~\eqref{eq:DR-step1}, we find the imaginary parts on the left and right sides of the equation cancel, leaving us with 
\begin{align}
\Re S(E) = \frac{1}{\pi} \mathcal{P} \int_{0}^{\infty} \frac{\Im S(E')}{E' - E}\, \diff E'
+ \sum_j \frac{\text{Res}\, S(E_j)}{E_j-E} .
\end{align}
This is an integral relation between the real and imaginary parts of $S$, analogous to the Kramers--Kronig relations~\cite{kramers1927diffusion,deL.Kronig:26} for real and imaginary parts of generalized indices of refraction in electrodynamics (corresponding to dispersion and absorption), which is the reason for the term \textit{dispersion relations}.

Let us now take care of the caveat above:
the condition $S(E) \to 0$ as $|E| \to \infty$ may not always hold.  In that case, consider instead
\begin{align}
F(E) = \frac{S(E) - S(E_0)}{E - E_0},
\end{align}
which falls off faster. Then
\begin{align}
F(E) = \frac{1}{\pi} \int_{0}^{\infty} \frac{\Im F(E')}{E' - E}\, \diff E' + \text{(bound-state contributions)} .
\end{align}
Resubstituting $F(E)$, we obtain
\begin{align}
S(E) = S(E_0) + \frac{E - E_0}{\pi} \int_{0}^{\infty} \frac{\Im S(E')}{(E' - E_0)(E' - E)}\, \diff E'
+ \text{(bound-state contributions)} ,
\end{align}
where $S(E_0)$ is the \textit{subtraction constant}, and $E_0$ the subtraction point.  
This is a once-subtracted dispersion relation.
If necessary, repeat the subtraction procedure $n$ times until convergence is guaranteed, with the result
\begin{align}
S(E) &= S(E_0) + S'(E_0)(E - E_0) + \dots  + \frac{(E - E_0)^n}{\pi} \int_{0}^{\infty} \frac{\Im S(E')}{(E' - E_0)^n (E' - E)}\, \diff E' + \text{(bound-state contributions)}.
\end{align}
Thus we can always ensure convergence of the integral, as long as the imaginary part does not grow faster than polynomially.
The price to pay is the need for $n$ subtraction constants, which have to be determined from additional input, besides the discontinuity of $S(E)$.

\section{Dispersion relations for Feynman diagrams}\label{sec:Feynman-diagrams}

So far, dispersion relations look more like a consistency check: real and imaginary parts of $S$-matrix elements are related to each other in a non-trivial way.  How can we turn this relation into a constructive or even predictive tool?  The answer is that, often, the imaginary parts of amplitudes can be calculated much more easily than the full amplitude itself.  We will first demonstrate this for (perturbative) Feynman diagrams, using \textit{Cutkosky rules}~\cite{Cutkosky:1960sp}; see, e.g., the textbooks Refs.~\cite{Itzykson:1980rh,Peskin:1995ev} for introductions. 
We now turn from non-relativistic quantum mechanics to relativistic quantum field theory; for the moment, this only means we will write dispersion relations in terms of the Mandelstam variable $s = E_{\text{CMS}}^2$ (in the center-of-mass [CMS] system) instead of $E$ or $k^2$. We will touch upon complications due to crossing symmetry later.  

Consider the simple Feynman loop integral, corresponding to a two-propagator, one-loop diagram
\begin{align}
M(s) = \frac{1}{2i}\int \frac{\diff^4 l}{(2\pi)^4}\,
\frac{1}{\bigl[( p_s - l)^2 - M^2 + i\epsilon\bigr] \big(l^2 - M^2 + i\epsilon\big)} , \label{eq:one-loop}
\end{align}
where $s=p_s^2$. 
How do we calculate $\disc M(s)$ or, equivalently, $\Im M(s)$?
Inspection shows that the only way to generate an imaginary part in Eq.~\eqref{eq:one-loop} is to actually hit the poles in the propagators, such that the $+i\epsilon$ prescription becomes relevant; this corresponds to setting the propagators \textit{on-shell}.
The precise derivation~\cite{Peskin:1995ev} shows that the diagram's discontinuity across the cut is obtained by replacing the (cut) propagators according to the prescription 
\begin{align}
\frac{1}{p^2 - M^2 + i\epsilon} \quad \longrightarrow\quad -2\pi i \,\delta(p^2 - M^2).
\end{align}
Comparing to Eq.~\eqref{eq:SokhWeier}, the factor of $2$ in front of the $\delta$-function can actually be understood from the fact that we calculate the discontinuity, i.e., the difference of the function above and below the cut.
We perform the resulting calculation for $M(s)$ in practice:
\begin{align}
\disc M(s) 
&= \frac{1}{2i}\int \frac{\diff^4 l}{(2\pi)^4}\, (-2\pi i)\,\delta\left(l^2-M^2\right)\,(-2\pi i)\,\delta\left((p_s-l)^2-M^2\right) = \frac{1}{2}\,\frac{i}{4\pi}\int \frac{|\vec{l}|^2\,\diff |\vec{l}| \,\diff \Omega_l}{2l^0}\,\delta\left(s-2\sqrt{s}\,l^0\right) \notag \\
&= \frac{1}{2}\,\frac{i}{8\pi}\int \sqrt{(l^0)^2-M^2}\,\diff l^0\,\diff \Omega_l\,\delta\left(s-2\sqrt{s}\,l^0\right) 
= \frac{1}{2}\,\frac{i}{8\pi}\,\frac{\sqrt{s/4 - M^2}}{2\sqrt{s}}\,\int \diff\Omega_l = \frac{1}{2}\,\frac{i}{8\pi}\,\sqrt{1 - \frac{4M^2}{s}} , \label{eq:discM}
\end{align}
where we have, step by step, used $\diff^4l = \diff l^0\, |\vec{l}|^2 \diff|\vec{l}| \,\diff\Omega_l$, simplified the argument of the second $\delta$-function according to $(p_s-l)^2-M^2 \to s-2\sqrt{s}\, l^0$ for $l^2=M^2$ in the CMS system, and used $|\vec{l}|\,\diff|\vec{l}| = l^0 \diff l^0$.  Defining the phase space factor $\sigma(s) = \sqrt{1-4M^2/s}$, we thus arrive at
\begin{align}
\Im M(s)
= \frac{1}{2}\cdot \frac{\sigma(s)}{16\pi} .
\end{align}

As the next step, we can use a dispersion relation to reconstruct $M(s)$ from its imaginary part.  
Since $\lim_{s\to\infty} \Im M(s) = \text{const.}$, we need a subtracted dispersion relation to make it converge.  Performing the substitutions $\zeta \equiv \sigma(z)$, $\sigma \equiv \sigma(s)$, we find
\begin{align}
M(s) &= M(0) + \frac{s}{\pi}\int_{4M^2}^\infty \frac{\Im M(z)\,\diff z}{z(z-s)} 
= M(0) + \frac{s}{32\pi^2}\int_{4M^2}^\infty \frac{\diff z}{z(z-s)}\sqrt{1-\frac{4M^2}{z}} \notag \\
&= M(0) + \frac{1}{16\pi^2}\int_0^1 \frac{\zeta^2}{\zeta^2-\sigma^2}\,\diff \zeta = M(0) + \frac{1}{16\pi^2}\times
\begin{cases}
1 - \frac{\sigma}{2}\log\frac{\sigma+1}{\sigma-1}, & s<0, \\
1 - \frac{1}{\sqrt{-\sigma^2}} \arctan \frac{1}{\sqrt{-\sigma^2}}, & 0<s<4M^2, \\
1 - \frac{\sigma}{2}\log\frac{1+\sigma}{1-\sigma} + i\frac{\pi}{2}\sigma, & s>4M^2.
\end{cases}
\end{align}
If you have ever calculated this loop integral in dimensional regularization, you will know that $\int \diff^4l/[(l^2\ldots)(l^2\ldots)]$ is (logarithmically) divergent. The above representation demonstrates that this divergence resides in the subtraction constant $M(0)$.

\section{Beyond Feynman diagrams: the use of unitarity}\label{sec:PionFF}
\subparagraph{Pion vector form factor}

Instructive as the previous example may be, the true strengths of dispersion relations may rightly be claimed to only come to light in non-perturbative contexts, such as the interactions of hadrons at low energies.  The elementary example we will discuss in some detail is the pion vector form factor, which is defined according to
\begin{align}
\langle \pi^+(p') \pi^-(p) | j_\mu(0) | 0 \rangle 
= (p' - p)_\mu \, F_\pi^V(s), \qquad
s = (p'+p)^2, \qquad \text{where} \quad
j_\mu = \tfrac{1}{2} \big( \bar{u}\gamma_\mu u - \bar{d}\gamma_\mu d \big) \label{eq:pionVFF}
\end{align}
is the isovector part of the electromagnetic current.   
The isoscalar part does not couple to two pions in the limit of $G$-parity conservation, where $G$-parity is related to charge conjugation $C$ and a rotation in isospin space according to $G = C e^{i\pi I_2}$.  This form factor can be measured in $e^+e^-\to\pi^+\pi^-$ (in the $s$-channel) or electron--charged-pion scattering (in the $t$-channel), disregarding minor isospin-breaking corrections.
$F_\pi^V(s)$ is a measure of the compositeness of the pion:
a point-like pion would correspond to $F_\pi^V(s) \equiv \text{const}$.  
It can be interpreted as the Fourier transform of the spatial charge distribution in a certain reference frame (the so-called Breit frame). 
Gauge invariance (or current conservation) dictates that there is only one single form factor, since the pion has no spin (unlike the nucleon with its electric and magnetic form factors);
there is no term $\propto (p'+p)_\mu$ in Eq.~\eqref{eq:pionVFF}.
The form factor normalization is given by
$F_\pi^V(s=0) = 1$, the pion's charge (in units of $e$).

The reader may rightly ask why we prefer to discuss a \textit{form factor} all of a sudden, hence the \textit{production} of a hadron pair from an electroweak current, having talked about \textit{scattering} in the non-relativistic context before.  The answer is that we still defer the complications of crossing symmetry: in contrast to a scattering amplitude that depends at least on CMS energy and scattering angle (or, equivalently, on two independent Mandelstam variables), the form factor is a function of the single variable $s$, which makes the following simpler.

\begin{figure}[t]
	\centering
	\includegraphics[width=.35\textwidth]{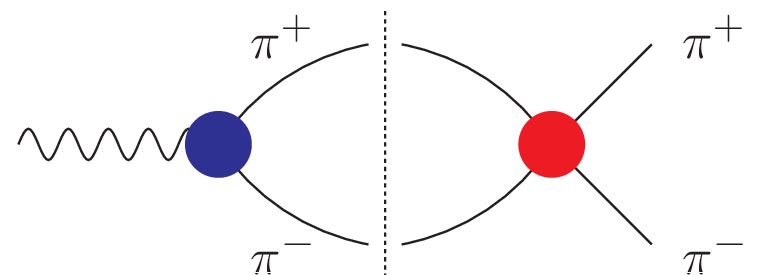}
	\caption{Unitarity relation for the pion vector form factor.  The curly line stands for the electromagnetic current, the blue disk denotes the form factor itself, while the red one stands for the pion--pion scattering amplitude.  The dotted line signifies the cut.}
	\label{fig:PionVFF}
\end{figure}
We begin by calculating the discontinuity of $F_\pi^V(s)$ due to $\pi\pi$ intermediate states, cf.\ Fig.~\ref{fig:PionVFF}, making maximal use of the result in the previous section.  We immediately find 
(removing a factor of $1/2$ compared to Eq.~\eqref{eq:discM} due to non-identical pions)
\begin{align}
(p'-p)_\mu \; \disc F_\pi^V(s) &= \frac{i}{32\pi^2} \, \sigma_\pi(s) \, F_\pi^V(s) \int \diff\Omega_l \, T^*_{\pi\pi}(s,z_l) (2l - p_s)_\mu , \quad \text{where} \quad 
p_s = p+p' , \quad p_s^2 = s , \quad
\sigma_\pi(s) = \sqrt{1 - \frac{4M_\pi^2}{s}} .
\end{align}
To calculate the angular integral with the vector $(2l - p_s)_\mu$,  we use the ansatz
\begin{align}
\int \diff\Omega_\ell \, T^*_{\pi\pi}(s,z_l) (2l - p - p')_\mu = L_1(s) (p+p')_\mu + L_2(s) (p'-p)_\mu , \label{eq:angular-ansatz}
\end{align}
as well as CMS kinematics
\begin{align}
l^0 = p^0 = p'^0 = \frac{\sqrt{s}}{2}, \qquad 
|\vec{l}| = |\vec{p}| = |\vec{p'}| = \frac{\sqrt{s}\, \sigma_\pi(s)}{2}, \qquad
z_l = \cos\theta_{\vec{p'}\vec{l}}, \qquad  
p^{(\prime)}l = \frac{s}{4} \left(1\pm\sigma_\pi^2(s) z_l\right) .
\end{align}
If we contract Eq.~\eqref{eq:angular-ansatz} with $p_s^\mu = (p+p')^\mu$, we find $0$ on the left-hand side due to $2l(p+p') - (p+p')^2=s-s=0$, which yields $L_1(s)=0$.  On the other hand, contracting with $(p-p')^\mu$ results in
\begin{align}
\int \diff\Omega_\ell \, T^*_{\pi\pi}(s,z_l) \, 2l (p-p') &= \int \diff\Omega_l \, T^*_{\pi\pi}(s,z_l) \, z_l \, \left(s-4M_\pi^2\right) 
= \left(s-4M_\pi^2\right) L_2(s),
\end{align}
and hence
\begin{align}
\int \diff \Omega_l \, T^*_{\pi\pi}(s,z_\ell) (2\ell - p_s)_\mu &= 2\pi\, (p'-p)_\mu  \int_{-1}^1 \diff z \, z \, T^*_{\pi\pi}(s,z) . \label{eq:disc-intermediate}
\end{align}
Recall the partial-wave expansion and the orthogonality of the Legendre polynomials,
\begin{align}
T_{\pi\pi}(s,z) &= 16\pi \sum_\ell (2\ell+1) P_\ell(z) \, t_\ell(s), \qquad
\int_{-1}^1 \diff z \, P_\ell(z) P_{\ell'}(z) = \frac{2\delta_{\ell\ell'}}{2\ell+1}, \qquad \text{thus} \quad
t_\ell(s) = \frac{1}{32\pi} \int_{-1}^1 \diff z \, P_\ell(z) \, T_{\pi\pi}(s,z) ;
\label{eq:pipi-pwe}
\end{align}
then we note (with $P_1(z) = z$) that Eq.~\eqref{eq:disc-intermediate} projects out the $P$-wave of the pion--pion scattering amplitude. The result for the discontinuity equation is therefore
\begin{align}
\frac{1}{2i} \disc F_\pi^V(s) = \Im \, F_\pi^V(s) = \sigma_\pi(s) \, t_1^*(s) \, F_\pi^V(s) \Theta(s-4M_\pi^2) = F_\pi^V(s) \, \sin \delta_1(s) \, e^{-i\delta_1(s)} \Theta(s-4M_\pi^2) , \label{eq:piVFF-unitarity}
\end{align}
where in the last step, we have expressed the (elastic) partial wave in terms of the phase shift $\delta_1(s)$, 
$t_\ell(s) =  \sin\delta_\ell(s) \, e^{i\delta_\ell(s)}/\sigma_\pi(s)$,
reflecting its relation to the $S$-matrix element $S_\ell(s) = 1 + 2i\sigma_\pi(s) t_\ell(s)$.

A first interesting conclusion we can draw from Eq.~\eqref{eq:piVFF-unitarity} is Watson's final-state theorem~\cite{Watson:1954uc}: the phase of the form factor needs to compensate for the phase of the partial wave,
\begin{align}
\text{arg}\, F_\pi^V(s) = \delta_1(s) .
\end{align}
Unitarity demands that the phase of the form factor below inelastic thresholds be equal to the $\pi\pi$ $P$-wave scattering phase shift.

\subparagraph{Omnès solution}

We keep on neglecting inelastic contributions and try to find a solution to the discontinuity equation~\eqref{eq:piVFF-unitarity}. Perturbative quantum chromodynamics suggests that, for asymptotically high energies, the pion form factor ought to drop $\asymp 1/s$~\cite{Chernyak:1977as,Farrar:1979aw,Efremov:1979qk,Lepage:1979zb,Lepage:1980fj}, which suggests it fulfills an unsubtracted dispersion relation:
\begin{align}
F_\pi^V(s) = \frac{1}{\pi} \int_{4M_\pi^2}^\infty \frac{\diff s'}{s'-s} \, \Im F_\pi^V(s'), \qquad
\Im F_\pi^V(s) = F_\pi^V(s) \, \sin \delta_1(s) \, e^{-i\delta_1(s)},
\end{align}
which thus forms an integral equation for $F_\pi^V(s)$.
A solution can be constructed analytically: consider a special solution $\Omega(s)$ that is free of zeros~\cite{Leutwyler:2002hm,Ananthanarayan:2011xt,RuizArriola:2024gwb,Leplumey:2025kvv} and normalized according to $\Omega(0)=1$.
From $\Omega(s+i\epsilon) = |\Omega(s)| e^{i\delta_1(s)}$ and the Schwarz reflection principle $\Omega(s-i\epsilon) = |\Omega(s)| e^{-i\delta_1(s)} = \Omega(s+i\epsilon) e^{-2i\delta_1(s)}$, we can conclude
\begin{align}
\log \Omega(s-i\epsilon) &= \log \Omega(s+i\epsilon) - 2i\delta_1(s) \qquad \text{and} \qquad 
\disc \log \Omega(s) = 2i\delta_1(s).
\end{align}
Suppose that the phase shift approaches a constant asymptotically, then we can write a once-subtracted dispersion relation for $\log \Omega(s)$,
\begin{align}
\log \Omega(s) &= \log \Omega(0) + \frac{s}{\pi} \int_{4M_\pi^2}^\infty \diff s' \frac{\delta_1(s')}{s'(s'-s)}, \quad \text{and therefore} \quad
\Omega(s) = \exp \left\{ \frac{s}{\pi} \int_{4M_\pi^2}^\infty \diff s' \frac{\delta_1(s')}{s'(s'-s)} \right\}. \label{eq:Omnes}
\end{align}
This is the Omnès solution~\cite{Omnes:1958hv}.  Its physical interpretation is that it resums $\pi\pi$ rescattering in $e^+e^- \to \pi^+\pi^-$ to all orders.

In reality, inelastic contributions to the discontinuity and hence the form factor dispersion relation ($4\pi$, $\pi\omega$, $\bar{K}K$\ldots) are not entirely negligible, but suppressed at low energies.
E.g., consider the pion charge radius, defined from the low-energy expansion of the form factor,
\begin{align}
F_\pi^V(s) = 1 + \frac{1}{6} \langle (r_\pi^V)^2 \rangle s + \Order(s^2), \qquad
\langle (r_\pi^V)^2 \rangle = 6 \left. \frac{\diff F_\pi^V(s)}{\diff s} \right|_{s=0}.
\end{align}
From the Omnès representation $F_\pi^V(s) = \Omega(s)$, we can read off a so-called sum rule for the radius by Taylor expansion,
\begin{align}
\langle (r_\pi^V)^2 \rangle_{\text{el.}} = \frac{6}{\pi} \int_{4M_\pi^2}^\infty \diff s' \frac{\delta_1(s')}{s'^2} = 0.419 \, \text{fm}^2,
\end{align}
which can be compared to $\langle (r_\pi^V)^2 \rangle_{\text{exp.}} = 0.429(4) \, \text{fm}^2$~\cite{Colangelo:2018mtw}.  The increase of the squared radius by a couple of percent can be attributed to inelastic effects.

In fact, if $F_\pi^V(s)=\Omega(s)$ is a solution of Eq.~\eqref{eq:piVFF-unitarity}, then so is any $P(s)\times \Omega(s)$, with $P(s)$ an arbitrary real polynomial.  
Data for $F_\pi^V(s)$ below $s=1\,\text{GeV}^2$ can be described well by the very simple representation $P(s)\times\Omega(s)$, $P(s)=1+\alpha s$, with a small $\alpha$ subsuming inelastic contributions~\cite{Hanhart:2016pcd}.
However, the requirement of $F_\pi^V(s) \asymp 1/s$ restricts the degree of the polynomial, as we will see below. 

$\Omega(s)$ is the central object for the dispersion-theoretic description of elastic final-state interactions between two hadrons only, for which there are many other applications.
E.g., the \textit{scalar} form factor of the pion uses the isoscalar scalar scattering phase shift $\delta_0^0(s)$ as input; it is theoretically highly interesting~\cite{Donoghue:1990xh,Moussallam:1999aq,Descotes-Genon:2000byu,Hoferichter:2012wf,Daub:2012mu,Celis:2013xja,Daub:2015xja,Winkler:2018qyg}, though not as directly measurable as the vector form factor.
Going beyond pions, $K_{l3}$ decays, $K^+ \to \pi^0 \ell^+ \nu_\ell$, are described in terms of two form factors, a scalar ($\delta_0^{1/2}$) and a vector ($\delta_1^{1/2}$) one, related in an analogous fashion to the corresponding $S$- and $P$-wave pion--kaon phase shifts~\cite{Bernard:2006gy,Bernard:2009zm,KTeV:2009moj}.
An interesting curiosity is the tensor form factor of the pion, defined as the matrix element of the antisymmetric quark tensor bilinear,
\begin{align}
\langle \pi^+(p') \pi^-(p) | \bar{q} \, \sigma_{\mu\nu} q | 0 \rangle = \frac{i}{M_\pi} (p_\mu p'_\nu - p'_\mu p_\nu) B_T^{\pi,q}(s),
\end{align}
where $\sigma_{\mu\nu} = \tfrac{i}{2}[\gamma_\mu,\gamma_\nu]$. This form factor is interesting for hadronic matrix elements of potential beyond-the-Standard-Model interactions, and it can be shown that~\cite{Cirigliano:2017tqn,Hoferichter:2018zwu}
\begin{align}
\Im B_T^{\pi,q}(s) = \sigma_\pi(s)\, t_1^*(s) B_T^{\pi,q}(s).
\end{align}
Thus, the corresponding form factor obeys exactly the same unitarity relation as $F_\pi^V$.
Assuming similar asymptotics $B_T^{\pi,q}(s\to\infty) \asymp 1/s$, we find
\begin{align}
B_T^{\pi,q}(s) = B_T^{\pi,q}(0) \times F_\pi^V(s) 
\end{align}
up to inelastic corrections,
and the $s$-dependence of the tensor form factor is strongly constrained by unitarity.

\begin{BoxTypeA}{Exercises}

To familiarize themselves with the integral representation of the Omnès function, the readers are invited to attempt the following simple exercises.
\begin{enumerate}
\item Demonstrate from the explicit representation~\eqref{eq:Omnes} that indeed $\arg\,\Omega(s) = \delta(s)$. 
\item Assume that the phase shift is a constant above a certain energy $s_0$, $\delta(s>s_0) = c\pi = \text{const}$.  Show  that asymptotically, $\Omega(s) \asymp s^{-c}$. An Omnès function that is to behave according to  $\Omega(s) \asymp 1/s$ therefore requires a phase shift input that approaches $\pi$ asymptotically.
\item Assume the phase shift of an infinitely narrow resonance, $\delta(s) = \pi \, \Theta(s-M^2)$. 
What is the resulting Omnès function?
\end{enumerate}
\end{BoxTypeA}

\section{Resonance poles}\label{sec:resonances}

We have understood in Sec.~\ref{sec:NRscattering} that analyticity severely limits the possibility of $S$-matrix singularities on the physical (or first) Riemann sheet: they can only occur on the real axis, as branch cuts corresponding to multi-particle intermediate states, or bound-state poles below threshold.  However, other singularities, in particular resonance poles, may be found on unphysical Riemann sheets.  We therefore want to understand how to continue scattering partial waves and form factors to unphysical sheets; for simplicity, we restrict ourselves to elastic pion--pion scattering and a two-sheet situation.

The $S$-matrix is unitary as a consequence of probability conservation,
$S^\dagger S = \mathbb{I}$, which implies 
$S_\ell^*(s)\, S_\ell(s) = 1$
for scattering states of definite angular momentum $\ell$.  
In the conventions above, the relation to $\pi\pi$ partial waves is given by
\begin{align}
S_\ell(s) = 1 + 2 i \sigma_\pi(s) \, t_\ell(s),
\end{align}
and it is an easy exercise to show that, as a result, we find the following unitarity relation for partial waves:
\begin{align}
2i\,\Im t_\ell(s) = \disc t_\ell(s) = t_\ell(s) - t_\ell(s^*) = 
t_\ell(s) - t_\ell^*(s) = 2 i \sigma_\pi(s) \, t_\ell^*(s)\, t_\ell(s), \label{eq:t-unitarity}
\end{align}
where we have made use of the Schwarz reflection principle again.
We could also have derived this ``by hand'' using Cutkosky rules, much as we did for the pion form factor in Sec.~\ref{sec:PionFF}.
Equation~\eqref{eq:t-unitarity} (which is the partial-wave analog of the optical theorem well-known from quantum mechanics) is more restrictive than the corresponding unitarity relation~\eqref{eq:piVFF-unitarity} for the form factor, as it is a non-linear relation that does not allow for any polynomial ambiguities.

Along the cut, first and second sheets are connected according to
\begin{align}
t_\text{I}(s+i\epsilon) = t_\text{II}(s-i\epsilon) = t_\text{I}^*(s-i\epsilon) 
\qquad \text{and} \qquad
t_\text{I}(s-i\epsilon) = t_\text{II}(s+i\epsilon) = t_\text{I}^*(s+i\epsilon).
\end{align}
We can therefore rewrite the discontinuity equation~\eqref{eq:t-unitarity} according to
\begin{align}
\disc t_\text{I}(s) &= t_\text{I}(s+i\epsilon) - t_\text{II}(s+i\epsilon) = 2i\sigma_\pi(s) \, t_\text{I}(s+i\epsilon)\, t_\text{II}(s+i\epsilon),
\end{align}
and solve for the second-sheet partial wave
\begin{align}
t_\text{II}(s) = \frac{t_\text{I}(s)}{1 + 2i\sigma_\pi(s)\, t_\text{I}(s)}.
\end{align}
For the continuation into the complex plane, it is often simpler to use 
$
\bar\sigma_\pi = \sqrt{{4M_\pi^2}/{s} - 1}
$
instead of $\sigma_\pi(s)$~\cite{Moussallam:2011zg,Hoferichter:2017ftn}, where $\bar\sigma_\pi(s \pm i\epsilon) = \mp i \sigma_\pi(s)$ selects the proper branch of the square root.  Thus
\begin{align}
t_\text{II}(s) = \frac{t_\text{I}(s)}{1 - 2 \bar\sigma_\pi(s) \, t_\text{I}(s)}. 
\label{eq:tII}
\end{align}

\begin{BoxTypeA}{Exercise}

Check that you arrive at the same result for $t_\text{II}$ starting from $S_\text{II}(s) = 1 - 2 i \sigma_\pi \, t_\text{II}(s)$
and using $S_\text{II}(s) = {1}/{S_\text{I}(s)}$.
    \end{BoxTypeA}

A pole in $t_\text{II}$ can occur only if
$1 - 2 \bar\sigma_\pi(s)\, t_\text{I}(s) = 0$, as $t_\text{I}(s)$ has no poles.
As
$1 - 2 \bar\sigma_\pi(s)\, t_\text{I}(s) = 1 + 2 i \sigma_\pi(s)\, t_\text{I}(s) = S_\text{I}(s)$, 
we recover the finding that $S$-matrix poles on the second sheet correspond to $S$-matrix zeros on the first sheet.  The condition for a resonance pole to occur at $s_\text{res}$ on the second sheet can therefore be formulated as a condition on the first sheet according to
\begin{align}
t_\text{I}(s_{\text{res}}) = \frac{1}{2 \bar\sigma(s_{\text{res}})} = - \frac{i}{\sigma(s_{\text{res}})}.
\end{align}

Now we can use the discontinuity relation for the form factor, Eq.~\eqref{eq:piVFF-unitarity}, to continue $F_\pi^V(s) \equiv F(s)$ to the second sheet in much the same way:
\begin{align}
F_\text{I}(s+i\epsilon) - F_\text{II}(s+i\epsilon) = - 2 \bar\sigma_\pi(s)\, F_\text{I}(s+i\epsilon) \, t_\text{II}(s+i\epsilon), \qquad \text{hence}
 \qquad 
F_\text{II}(s) = \big(1 + 2 \bar\sigma_\pi(s)\, t_\text{II}(s)\big) F_\text{I}(s)
= \frac{F_\text{I}(s)}{1 - 2 \bar\sigma_\pi(s)\, t_\text{I}(s)}.
\end{align}
The denominator is the same as in $t_\text{II}(s)$, which proves the universality of resonance pole positions in scattering and production amplitudes:
e.g., the pole position of the $\rho(770)$ in the $\pi\pi$ $P$-wave is identical to the one in $F_\pi^V(s)$.

The difficult part is the pole determination in the scattering partial wave~\cite{Caprini:2005zr,Garcia-Martin:2011nna}, and we will minimally comment on the dispersion-theoretical techniques to achieve this below in Sec.~\ref{sec:scattering}. 
Near the $\rho$ pole, we can approximate the $P$-wave amplitude according to
\begin{align}
t_{1,\text{II}}(s) \simeq \frac{g_{\rho\pi\pi}^2}{48\pi} \, \frac{s - 4M_\pi^2}{s_\rho - s}, \qquad 
s_\rho = \left(M_\rho - \frac{i}{2}\Gamma_\rho\right)^2, \label{eq:F-II}
\end{align}
where we have identified mass and width of the $\rho$ resonance from real and imaginary part of $\sqrt{s_\rho}$.
This defines a (complex) $\rho\to\pi\pi$ coupling constant $g_{\rho\pi\pi}$ from the residue of the pole.   
In practice $|g_{\rho\pi\pi}| \simeq 6.0(1)$, $\arg(g_{\rho\pi\pi}) \simeq -5(1)^\circ$~\cite{Hoferichter:2023mgy}, and the corresponding coupling is almost real; this need not always be true, cf., e.g., the scalar $f_0(500)=\sigma$ resonance, for which $\arg(g_{\sigma\pi\pi}) \simeq -80 (6)^\circ$)~\cite{Hoferichter:2023mgy}.

But now, the determination of form factor residues is simple: 
again near the $\rho$ pole, we can approximate
\begin{align}
F_{\pi,\text{II}}^V(s) \simeq g_{\rho\gamma} \, g_{\rho\pi\pi} \, \frac{s_\rho}{s_\rho - s},
\end{align}
and equating the poles on both sides of the form factor discontinuity equation~\eqref{eq:F-II} yields
\begin{align}
g_{\rho\gamma}\, g_{\rho\pi\pi} \, \frac{s_\rho}{s_\rho - s} = 2 i \sigma_\pi(s_\rho) \, F_{\pi,\text{I}}^V(s_\rho)\, \frac{g_{\rho\pi\pi}^2}{48\pi}\, \frac{s_\rho - 4M_\pi^2}{s_\rho - s} \qquad \text{and} \qquad
\frac{g_{\rho\gamma}}{g_{\rho\pi\pi}} = i \frac{\sigma_\pi^3(s_\rho)}{24\pi} \, F_{\pi,\text{I}}^V(s_\rho).
\end{align}
We therefore only need to evaluate $F_{\pi,\text{I}}^V(s) \propto \Omega(s)$ on the first sheet to extract the residue relative to the one in the pion--pion partial wave.  Access to both first and second sheet of the form factor allows one to even extract $g_{\rho\gamma}$ and $g_{\rho\pi\pi}$ independently~\cite{Kirk:2024oyl}.

\section{Some remarks on dispersion relations for scattering processes}\label{sec:scattering}

We have so far considered dispersion relations for a single variable; for form factors, naturally, and in the context of scattering amplitudes by concentrating on partial waves.  However, the fact that scattering amplitudes are actually functions of \textit{two} independent Mandelstam variables, plus the fact that relativistic quantum field theory leads to crossing symmetry, has important consequences on the analytic structure of partial waves beyond what we have considered so far.

For a scattering reaction with momentum flow $p_1 \, p_2 \to p_3 \, p_4$, the conventional Mandelstam variables are given by
$s = (p_1 + p_2)^2 = (p_3 + p_4)^2$, 
$t = (p_1 - p_3)^2 = (p_2 - p_4)^2$, and 
$u = (p_1 - p_4)^2 = (p_2 - p_3)^2$,
subject to the on-shell constraint $s + t + u = \sum_{i=1}^4 M_i^2$.
Analyticity and crossing symmetry dictate that there is a single analytic function $T(s,t)$ that describes the scattering reaction and all its crossed channels in the whole Mandelstam plane, including all three physical regions. 
This conjecture can be justified from
knowledge on non-relativistic quantum mechanical scattering,  
axiomatic quantum field theory,  
and the study of (perturbative) Feynman diagrams.
Thus $T$ is analytic in two variables.

Cuts above threshold---$s_\text{thr} = 4M_\pi^2$ for our simple example of pion--pion scattering---occur not only in the variable $s$, but in all three variables.\footnote{We neglect the possibility of bound-state poles in the $T$-matrix, which is irrelevant for $\pi\pi$ scattering at physical pion masses.}  
One possible path forward is the use of double-variable dispersion relations (so-called Mandelstam representations~\cite{Mandelstam:1958xc,Mandelstam:1959bc}), which is a rather advanced topic.  Probably more often, the problem is reduced to single-variable dispersion relations, which can be achieved, e.g., by one of the following two possibilities:
\begin{enumerate}
\item \textit{fixed-$t$ dispersion relations} (or similar); a special case are dispersion relations for $t=0$, which correspond to \textit{forward dispersion relations}.  
\item By integrating over the cosine of the scattering angle (or, equivalently, $t$), project onto partial waves to arrive at \textit{partial-wave dispersion relations}.
\end{enumerate}
We briefly discuss both of these options in turn.

\subparagraph{Fixed-\textit{t} dispersion relations}

We want to analyze the scattering amplitude for fixed value of $t$,
$T(s,t=t_0) \equiv f(s)$.  What cuts does $f(s)$ have?
Obviously, there is the physical or unitarity or right-hand cut for
$s \stackrel{\text{CMS}}{=} 4\left(M_\pi^2 + \vec{p}^2\right) \geq 4M_\pi^2$.
But crossing symmetry implies that there is another cut for  
$u = 4M_\pi^2 - t_0 - s \geq 4M_\pi^2$,
which leads to a cut for $s \leq -t_0$, extending to $s \to -\infty$.
For this reason, this second cut in $s$ that originates in the unitarity cut in the variable $u$ is mostly called a \textit{left-hand cut}.
As a result, a dispersion relation for $f(s)$ contains \textit{two} contributions,
\begin{align}
f(s) &= \frac{1}{\pi} \int_{4M_\pi^2}^\infty \frac{\Im f(s')}{s'-s} \, \diff s'
+ \frac{1}{\pi} \int_{-\infty}^{-t_0} \frac{\Im f(s')}{s'-s} \, \diff s' = \frac{1}{\pi} \int_{4M_\pi^2}^\infty \diff s' \left( \frac{1}{s'-s} + \frac{1}{s'-u} \right) \Im f(s'),
\end{align}
where the first and second terms correspond to right- and left-hand cuts in $s$, respectively, and the second, alternative, formulation keeps the implicit variable $u$ instead.

\subparagraph{Partial-wave dispersion relations}

We have already introduced the partial-wave projection for pion--pion scattering in Eq.~\eqref{eq:pipi-pwe},
\begin{align}
t_\ell(s) = \frac{1}{32\pi} \int_{-1}^1 \diff z \, P_\ell(z) T\big(s,t(s,z),u(s,z)\big) , \qquad \text{where} \quad
t(s,z) = -\frac{s-4M_\pi^2}{2}(1-z), \quad 
u(s,z) = -\frac{s-4M_\pi^2}{2}(1+z).
\end{align}
The maximum (positive) value for $t/u$ in the range $-1 \leq z \leq 1$ is $4M_\pi^2 - s$, hence the corresponding cuts $t,u \geq 4M_\pi^2$ are hit for all $s \leq 0$.  
As a consequence, the left-hand cut here extends to $s=0$, and the partial-wave dispersion relation reads
\begin{align}
t_\ell(s) &= \frac{1}{\pi} \int_{4M_\pi^2}^\infty \frac{\Im t_\ell(s')}{s'-s} \, \diff s'
+ \frac{1}{\pi} \int_{-\infty}^0 \frac{\Im t_\ell(s')}{s'-s} \, \diff s'.
\end{align}

\subparagraph{Roy equations for pion--pion scattering}

In practice, dispersion relations for scattering rarely work without subtractions. The Froissart--Martin bound~\cite{Froissart:1961ux,Martin:1962rt} suggests that two are safely sufficient.  
The problem is that in a fixed-$t$ dispersion relation, subtractions actually result in $t$-dependent subtraction \textit{functions},
\begin{align}
T(s,t) = C(t) + \frac{1}{\pi} \int_{4M_\pi^2}^\infty \diff s' \left[
\frac{s^2}{s'^2(s'-s)} + \frac{u^2}{s'^2(s'-u)}
\right] \Im T(s',t),
\end{align}
where we again concentrate on the example of $\pi\pi$ scattering, and for the sake of the argument, we omit any complications due to isospin.
Crossing symmetry forbids subtractions $\propto s-u$, and $s+u=4M_\pi^2 - t$ can be absorbed in $C(t)$.
The subtraction function $C(t)$ can be fixed from $s \leftrightarrow t$ crossing symmetry:
\begin{align}
T(0,t) &= C(t) + \frac{1}{\pi} \int_{4M_\pi^2}^\infty \diff s' \, 
\frac{(4M_\pi^2 - t)^2}{s'^2 (s' - 4M_\pi^2 + t)} \, \Im T(s',t) \notag\\
= T(t,0) &= C(0) + \frac{1}{\pi} \int_{4M_\pi^2}^\infty \diff s' 
\left( \frac{t^2}{s'^2(s'-t)} + \frac{(4M_\pi^2 - t)^2}{s'^2 (s'-4M_\pi^2+t)} \right) 
\Im\, T(s',0).
\end{align}
The remaining constant $C(0)$ is related to the amplitude at threshold, i.e., the scattering length:
\begin{align}
T\big(4M_\pi^2,0\big) = C(0) + \frac{1}{\pi} \int_{4M_\pi^2}^\infty \diff s' \, 
\frac{16 M_\pi^4}{s'^2 (s' - 4M_\pi^2)} \, \Im T(s',0).
\end{align}
Now, expand $\Im T(s,t)$ in partial waves and project $T(s,t)$ onto partial waves;
the result (including isospin $I$ again) can be cast into the generic form
\begin{align}
t_\ell^I(s) &= k_\ell^I(s) + \sum_{I'=0}^2 \sum_{\ell'=0}^\infty 
\int_{4M_\pi^2}^\infty \diff s' \, K_{\ell\ell'}^{II'}(s,s') \, \Im t_{\ell'}^{I'}(s'),
\end{align}
where $k_\ell^I(s)$ is the subtraction polynomial, depending only on the pion--pion $S$-wave scattering lengths, and $K_{\ell\ell'}^{II'}(s,s')$ are analytically known kernel functions, including the diagonal Cauchy kernel $\delta_{II'}\delta_{\ell\ell'}/(s'-s-i\epsilon)$ plus left-hand-cut contributions.  
Rewriting the partial waves according to
$t_\ell^I(s) = e^{i\delta_\ell^I(s)} \sin \delta_\ell^I(s)/\sigma_\pi(s)$,
this turns into a system of coupled integral equations for the $\pi\pi$ phase shifts $\delta_\ell^I(s)$.  These are the Roy equations~\cite{Roy:1971tc}, of which modern high-precision analyses exist~\cite{Ananthanarayan:2000ht,Garcia-Martin:2011iqs} (cf.\ also Ref.~\cite{Colangelo:2025sah} for a more extended pedagogical introduction to the subject), the main source of our knowledge of the corresponding phase shifts at low energies.  Similar partial-wave dispersion relations (although technically implemented slightly differently~\cite{Hite:1973pm}) have also been studied in detail for pion--kaon~\cite{Buettiker:2003pp,Pelaez:2020gnd}, pion--photon~\cite{Hoferichter:2011wk,Hoferichter:2019nlq}, and pion--nucleon scattering~\cite{Hoferichter:2015hva}.

\section{Dispersion relations for three-body decays}\label{sec:decays}

For a 4-point function with different masses and $M_1 > M_2 + M_3 + M_4$, a fourth physical region appears: in addition to the three scattering processes with their corresponding kinematics, there is the decay region. 
For illustration, we discuss the particularly simple example of the decay
$\omega(P) \to \pi^+(p_1)\pi^-(p_2)\pi^0(p_0)$, which displays a high degree of crossing symmetry and is very selective in terms of possible partial waves. 
The amplitude is once more described in terms of three Mandelstam variables, 
$s = (p_1+p_2)^2 = (P-p_0)^2$, 
$t = (p_0+p_2)^2 = (P-p_1)^2$, and
$u = (p_0+p_1)^2 = (P-p_2)^2$.
The decay region is constrained by 
\begin{align}
4M_\pi^2 \leq s, t, u \leq (M_\omega - M_\pi)^2 .
\end{align}
The decay $\omega\to3\pi$ is a process of odd intrinsic parity, which requires the amplitude to involve the Levi-Civita symbol and be of the form
\begin{align}
\mathcal{M}(s,t,u) = i \epsilon_{\mu\nu\alpha\beta}\, 
\epsilon^\mu(P) p_1^\nu p_2^\alpha p_0^\beta\, \F(s,t,u),
\end{align}
where $\epsilon^\mu(P)$ is the polarization vector of the $\omega$. 
The ($s$-channel) partial-wave expansion is complicated by the fact that the $\omega$ has spin, and reads~\cite{Jacob:1959at}
\begin{align}
\F(s,t,u) = \sum_{\ell \ \text{odd}} f_\ell(s) P'_\ell(z_s)
= f_1(s) + \ldots ,
\end{align}
where $P'_\ell$ denotes the derivatives of Legendre polynomials, and only odd partial waves contribute due to Bose symmetry.
Partial waves are obtained via the relation
\begin{align}
f_\ell(s) = \frac{1}{2} \int_{-1}^1 \diff z_s \big[ P_{\ell-1}(z_s) - P_{\ell+1}(z_s) \big] \F(s,t,u), \qquad \text{in particular} \quad
f_1(s) = \frac{3}{4} \int_{-1}^1 \diff z_s\,(1-z_s^2)\,\F(s,t,u).
\end{align}
It can be shown that, neglecting discontinuities in $F$-waves and higher, $\F(s,t,u)$ can be decomposed in terms of single-variable amplitudes according to~\cite{Niecknig:2012sj,Hoferichter:2012pm}
\begin{align}
\F(s,t,u) = \F(s) + \F(t) + \F(u), \label{eq:reconstruction}
\end{align}
where $\F(s)$ has only a right-hand cut.
Relations like Eq.~\eqref{eq:reconstruction} are often called \textit{reconstruction theorems}~\cite{Stern:1993rg,Knecht:1995tr,Zdrahal:2008bd}; in a very simplified way, they can be understood as symmetrized partial-wave expansions in all three variables.

We can derive a unitarity relation for $f_1$ based on $\pi\pi$ intermediate states only.  
Projecting out the $P$-wave, we find an Omnès-like unitarity relation along the right-hand cut,
\begin{align}
\disc f_1(s) = 2i\, f_1(s)\sin\delta_1(s)\, e^{-i\delta_1(s)}\Theta(s-4M_\pi^2).
\label{eq:disc-f1}
\end{align}
But 
\begin{align}
    f_1 = \F(s) + \hat\F(s) , \qquad 
    \hat{\F}(s) = 3 \langle (1-z^2)\F\rangle (s),
\qquad
\langle z^n f \rangle (s) = \frac{1}{2}\int_{-1}^1 \diff z\,z^n f(t(s,z)) \label{eq:angular-av}
\end{align}
has left-hand cuts, too, 
due to $\hat\F(s)$ that results from the partial-wave projection of $\F(t)+\F(u)$.
Hence there is no simple Omnès solution as for a form factor.

Along the right-hand cut, $\disc f_1(s) = \disc \F(s)$, hence we can rewrite Eq.~\eqref{eq:disc-f1} according to
\begin{align}
\disc \F(s) 
= 2i\big(\F(s)+\hat{\F}(s)\big)\sin\delta(s)e^{-i\delta(s)}\Theta(s-4M_\pi^2).
\label{eq:disc-F}
\end{align}
Equation~\eqref{eq:disc-F} is an inhomogeneous Omnès problem with inhomogeneity $\hat{\F}$.
To solve it, we observe that we already know the homogeneous solution with $\hat{\F}=0$: 
$\F(s) = P(s)\Omega(s)$, the Omnès solution with multiplicative polynomial ambiguity. The ansatz to proceed is therefore to consider the discontinuity of $\F(s)/\Omega(s)$, making use of $\Omega(s\pm i\epsilon) = |\Omega(s)|\, e^{\pm i\delta(s)}$:
\begin{align}
\disc\left(\frac{\F(s)}{\Omega(s)}\right) 
&= \frac{\F(s+i\epsilon)}{\Omega(s+i\epsilon)} - \frac{\F(s-i\epsilon)}{\Omega(s-i\epsilon)}
= \frac{\F(s+i\epsilon)e^{-i\delta(s)} - \F(s-i\epsilon)e^{i\delta(s)}}{|\Omega(s)|} 
= \frac{\disc \F(s) e^{i\delta(s)} - 2 i \,\F(s+i\epsilon) \sin\delta(s) }{|\Omega(s)|}  \notag\\
&= \frac{2i\,\hat{\F}(s)\,\sin\delta(s)}{|\Omega(s)|},
\end{align}
where we have employed Eq.~\eqref{eq:disc-F} in the last step.
Writing therefore a dispersion relation for $\F(s)/\Omega(s)$ and multiplying the result with the Omnès function, we arrive at 
\begin{align}
\F(s) = \Omega(s) \Bigg\{ P(s) + \frac{s^n}{\pi} \int_{4M_\pi^2}^\infty \frac{\diff s'}{s'^n}\,
\frac{\hat{\F}(s')\sin\delta(s')}{|\Omega(s')|(s'-s)} \Bigg\}, \label{eq:KT}
\end{align}
where $P(s)$ is a subtraction polynomial of degree $n-1$.  
We can interpret this result as the sum of a homogeneous part $\Omega(s) \times P(s)$ that denotes direct (two-body) rescattering in the $s$-channel, and the dispersion integral represents the more complicated interplay with the crossed-channel dynamics.

We find that the formal solution, known as Khuri--Treiman equations~\cite{Khuri:1960zz}, requires 
$\F(s)$ to be calculated from $\hat\F(s)$ by dispersive integration via Eq.~\eqref{eq:KT}, and 
$\hat{\F}(s)$ to be calculated from $\F(t)$ via angular integration, cf.\ Eq.~\eqref{eq:angular-av}.  This system can either be solved by iteration, or by discretization and matrix inversion~\cite{Niecknig:2015ija}.  
It involves several subtleties:
\begin{enumerate}
\item The discontinuity is complex in the decay region; the resulting single-variable amplitude $\F(s)$ therefore does not obey the Schwarz reflection principle. The correct amplitude is obtained by analytic continuation in the decay mass from the scattering to the decay region~\cite{Bronzan:1963mby}.
\item The angular integration $\hat{\F}$ has to be deformed into the complex plane in order to avoid the right-hand cut.
\item The discontinuity has a singularity at the pseudothreshold 
$s=(M_\omega - M_\pi)^2$, so dispersive integration is numerically difficult.
\end{enumerate}
We refer to, e.g., Ref.~\cite{Gasser:2018qtg} for a discussion of several of these subtleties.
The Khuri--Treiman formalism, originally devised to describe $K\to3\pi$ decays~\cite{Khuri:1960zz,Bernard:2024ioq}, has been revived with the advent of modern high-precision pion--pion scattering phase shifts, in particular for $\eta\to3\pi$ decays~\cite{Kambor:1995yc,Anisovich:1996tx,Colangelo:2018jxw,Akdag:2021efj}, but with a multitude of other modern applications~\cite{Albaladejo:2019huw,Niehus:2021iin,Stamen:2022eda,JPAC:2023nhq,Hoferichter:2025lcz} in the literature.

\section{Conclusions}
\label{sec:conclusions}

The purpose of this chapter was to provide a brief and largely pedagogical introduction to the principles underlying dispersion theory as rooted in causality and probability conservation, and demonstrate some applications relevant for the modern theory of hadron interactions at low-to-medium energies.  Many sources treat different aspects that we have merely superficially touched upon with much more coherence and rigor.  
References~\cite{Hannesdottir:2022bmo,Mizera:2023tfe}, e.g., discuss the physics of the analytic $S$-matrix in a lot more detail, as does the classic text of Ref.~\cite{Eden:1966dnq}.  A much more detailed overview of various applications in hadron physics is given, e.g., in Ref.~\cite{Oller:2019rej}.  We recommend all readers looking for more in-depth treatments of this topic to consult these references and further literature recommended therein.

\begin{ack}[Acknowledgments]%
The content of this chapter is based on lectures first held at the CRC Summer School on Low-Energy Strong Interactions in Beijing, China, in August 2019, and I am very grateful to the organizers of that school back then.  For parts of them, I have profited a lot from similar lectures given by Jos\'e R.\ Pel\'aez at the School on Amplitude Analysis in Bad Honnef, Germany, in August 2011, to whom I am deeply indebted.  
Finally, I would like to thank Martin Hoferichter and Barry Holstein for helpful comments on the manuscript.
\end{ack}

\bibliographystyle{elsarticle-num}
\bibliography{reference}

\end{document}